\documentclass{aa} 
\usepackage{natbib}
\usepackage{graphicx}
\usepackage{txfonts}
\usepackage{hyperref}
\begin{document} 

   \title{UV and NIR size of the low-mass field galaxies: the UV compact galaxies}
\author{Cheng Cheng\inst{1,2}, Cong Kevin Xu\inst{1,2}, 
     Lizhi Xie\inst{3}, Zhizheng Pan\inst{4}, Wei Du\inst{1}, Gustavo Orellana-Gonz\'alez\inst{6,5}, Xu Shao\inst{1,2}, Shumei Wu\inst{1,2}, Roger Leiton\inst{5}, Jia-Sheng Huang\inst{1,2}, Sophia Yu Dai\inst{1,2}, Paulina Assmann\inst{5},\and 
     Nicole Araneda\inst{5}
}
	\institute{Chinese Academy of Sciences South America Center for Astronomy, National Astronomical Observatories, CAS, Beijing 100101, China. \email{chengcheng@nao.cas.cn}
    \and
    China-Chile Joint centre for Astronomy, Camino El Observatorio 1515, Las Condes, Santiago, Chile.
        \and
        Tianjin Astrophysics centre, Tianjin Normal University, Tianjin 300387, People‚Äôs Republic of China.
        \and
    Purple Mountain Observatory, Chinese Academy of Sciences, 2 West-Beijing Road, Nanjing 210008, People's Republic of China.
        \and
    Departamento de Astronom\'ia, Universidad de Concepci\'on, Casilla 160-C, Concepci\'on, Chile.
        \and
        Instituto de F\'isica y Astronom\'ia, Universidad de Valpara\'iso, Avda. Gran Breta\~na 1111, Valpara\'iso, Chile. 
}
\titlerunning{The local UV compact galaxies}
\authorrunning{Cheng et al.}
\date{Received ...; accepted ...}


  \abstract
   {Most of the massive star-forming galaxies are found to have `inside-out' stellar mass growth modes, which means the inner parts of the galaxies mainly consist of the older stellar population, while the star forming in the outskirt of the galaxy is still ongoing.}
   {The high-resolution HST images from Hubble Deep UV Legacy Survey (HDUV) and Cosmic Assembly Near-infrared Deep Extragalactic Legacy Survey (CANDELS) 
projects with the unprecedented depth in both F275W and F160W bands are the perfect data sets to study the forming and formed stellar distribution directly.}
   {We selected
the low redshift ($0.05 < z_{\rm spec} < 0.3$) galaxy sample from the GOODS-North field where the HST F275W and F160W images are available. 
        Then we measured the half light radius in F275W and F160W bands, which are the indicators of the star formation and stellar mass.}
   {By comparing the F275W and F160W half light radius, we find the massive galaxies are mainly follow the `inside-out' growth mode, 
        which is consistent with the previous results. Moreover, the HST F275W and F160W images reveal that some of the low-mass galaxies ($<10^8M_\odot$) have 
        the `outside-in' growth mode: their images show a compact UV morphology, implying an ongoing star formation in the galaxy centre, while 
        the stars in the outskirts of the galaxies are already formed. The two modes transit smoothly at stellar mass range about $10^{8-9}M_\odot$ with a large scatter. 
        We also try to identify the possible neighbour massive galaxies from the SDSS data, which represent the massive galaxy sample. 
        We find that all of the spec-z selected galaxies have no massive galaxy nearby. Thus the `outside-in' mode we find in the
        low-mass galaxies are not likely originated from the environment.}
   {}


   \maketitle
%

\section{Introduction}

How and when the galaxies assemble the stellar mass are the key questions in the study of a galaxy's formation and evolution. 
Since the galaxy stellar mass assembly history is mainly the history of star formation, the study of the star-formation process
is crucially important in understanding a galaxy's properties. Previous observations of local and high-redshift galaxies show that 
the massive galaxies follow `inside-out' growth modes \citep{Nelson2012, Dokkum2013, Dokkum2014, Pan2015, Nelson2016, Lilly2016, Liu2016, Belfiore2017, Gobat2017, Tacchella2018, Nelson2019}.
However, little is known about the growth and quenching of low-mass field galaxies, because, on the one side, 
the deep surveys have small volumes for low redshifts, and, on the other side, the large area surveys 
\citep[e.g. ][]{Driver2011} are not deep enough to reach the stellar mass below 10$^8 M_\odot$.

The formation and evolution of the low-mass galaxy population might be different from the massive galaxies \citep{Faber2007}. 
Low-mass galaxies are typically located in dark matter halos with shallow gravitational potential, so they are more 
easily affected by the environment from outside \citep{French2016}, or by the supernova (SN) / 
active galactic nucleus (AGN) from inside   
\citep{MacLow1999, Silk2011, Silk2017, Dashyan2018}. Previous studies suggest that the star formation rate (SFR) in small galaxies 
is suppressed by the nearby massive galaxies \citep{Geha2012, Guo2017, Schaefer2017}. Nevertheless, because of 
their faintness and small size, we still lack knowledge regarding the formation path, the quenching, and 
star-forming gradients of the low-mass galaxy population. 

To study the formation mechanism of the low-mass galaxies, we need the data to show the spatially resolved 
star formation and stellar mass. The stellar mass can be represented by the rest-frame NIR images, since 
the largest stellar populations are mainly low-mass stars, while the spatially resolved star formation 
is not so easy to obtained. Popular star-formation observations such as the hydrogen recombination 
lines from the narrowband filter \citep[e.g. ][]{Hao2018} or Far-infrared (FIR) image \citep{Kennicutt2011} usually 
have a beam size that can only resolve the extent, massive galaxies, or the dwarf galaxies from the local group or nearby
galaxy clusters \citep{Kennicutt2003}, which may be affected by the environment. 
Integrated Field Unit (IFU) observations (e.g. The Multi Unit Spectroscopic Explorer (MUSE) or 
Spectrograph for INtegral Field Observations in the Near Infrared (SINFONI) data from Very Large Telescope (VLT) 
for low- and high-redshift galaxies with Adaptive Optics (AO), and the Mapping Nearby Galaxies at APO (MaNGA), 
the Calar Alto Legacy Integral Field Area Survey (CALIFA) data with spatial resolution about 1'') 
can provide spatial-resolved H$\alpha$ and H$\beta$ maps, optimal for studies of star-formation distributions in galaxies. However, these observations require long exposure times, resulting in low survey efficiency, 
especially for the population of low-mass galaxies. 
Interferometry sub-millimetre (e.g. Atacama Large Millimetre Array) with the high spatial resolution is also time-consuming and would be more 
suitable for studying the higher redshift galaxies \citep{Nelson2019}.

Previous studies have shown that the low-mass galaxies may have a `outside-in' growth mode \citep{Zhang2012} 
The Transition stellar mass of the galaxies from `outside-in' to `inside-out' is about $10^{10}M_{\odot}$\citep{Pan2015}. 
The MANGA and CALIFA results also reveal a flatter sSFR, metallicity, and stellar age 
distributions along the galaxy radius when the stellar mass is lower \citep{Perez2013, Belfiore2017}.
Moreover, detailed analyses from the MANGA data found that the strength and the concentration of the quenched 
area have a higher fraction of `inside-out quenching' mode in massive halo \citep{Lin2019}. \citet{Schaefer2017, Schaefer2019} 
compared the H$\alpha$ and the stellar continuum scale with SAMI IFU data, and found the fraction of the galaxies with the 
`centrally concentrated' star-formation core increases with the environment density, implying the environment quenching 
process may account for the SFR concentration.  

Besides the spectroscopy approach to resolving the star-formation distribution, 
to understand the relation between the star formation and stellar mass, one simple method is 
to compare the rest-frame ultraviolet (UV) and near-infrared (NIR) bands' images under a similar resolution. 
The UV image shows a hint of the star-formation morphology, while the NIR image can be used to 
trace the stellar mass distribution. Moreover, The low-mass galaxies might be too young to have much 
dust that transfers the UV energy to the FIR \citep{Sadavoy2019}. With less dust extinction, 
the direct comparison of the rest-frame UV and NIR images would be more efficient in studying the low-mass galaxy populations.

The recent Hubble Deep UV Legacy Survey \citep[HDUV][]{Oesch2018} project provides us a unique chance to 
study the galaxy's UV morphology in GOODS-South and North fields ($\sim$ 100 arcmin$^2$ in total). 
On the other hand, earlier HST survey project Cosmic Assembly Near-infrared Deep Extragalactic Legacy Survey (CANDELS) 
has also achieved great success in investigating galaxy properties \citep{Grogin2011, Koekemoer2011}. 
Taking advantage of the HST spatial resolution, for the low redshift galaxies, the F160W image from the CANDELS project 
can be used to trace the stellar mass 
distribution, while the F275W image from the HDUV project traces the SFR distribution. The high spatial resolution and 
the unprecedented survey depth from HDUV and CANDELS projects make it possible to study the relationship between 
the forming and formed stellar components in galaxies of stellar mass as low as $10^7 M_\sun$, and with sub-kpc 
resolution\footnote{The spatial resolutions of the UV and NIR image are about 0.1'' and 0.17'', which is about 
0.3, 0.5 and 0.7 kpc at redshift 0.1, 0.2 and 0.3, respectively. }.

Since the observations have shown that,
for most galaxies, the morphology of SFR distribution is not as smooth
as that of stellar mass distribution\citep{Paz2007,Kennicutt2008}, we
chose to use the non-parametric method in this work.  Namely, we
measured the half-light radii from the F275W and F160W images directly in
order to characterise the scales of both the SFR and stellar mass
distributions. The ratio between the F275W and F160W half-light radii shows
whether the star formation is more extent
than the stars. A similar comparison
to identify the `inside-out' mode has been applied to the IFU \citep{Schaefer2017} 
and the H$\alpha$ image from HST \citep{Nelson2012}.
In this paper, we make use of the recent HST F275W and F160W images and measure the half light radii in these bands, 
aiming to investigate the low redshift galaxies' formation path.
Throughout the paper, we assume a cosmological model, with $H_0 = 70\rm km/s/Mpc$, $\Omega_{\rm m}$ = 0.3, and $\Omega_{\Lambda}$ = 0.7.

\section{Sample selection}

We select our sample from the CANDELS catalogue in GOODS-North field \citep{Barro2019}
and the image data released by HDUV \footnote{https://archive.stsci.edu/prepds/hduv/}.
The CANDELS catalogue is selected from the F160W image deep to 27 AB mag (5$\sigma$, FWHM radius aperture), 
and the multi-wavelength catalogue is generated by the  Template FITting method \citep[TFIT][]{Laidler2007} 
, which deblends the low-resolution images
by the spatial information from the high-resolution image.

In what follows, we describe the details of the sample selection: 
\begin{itemize}  
\item{} The sample is in the GOODS-North field and is confined to
areas covered by deep UV exposures in the HDUV survey,
with the detection limits at F275W = 27.4 AB mag.
\item{} Its magnitude limited at F606W = 24 AB mag.
For galaxies of z=0.2, this corresponds approximately to a mass limit of 
$M_* = 10^7 M_\sun$ \citep[e.g. Fig. 2 in][]{Mahajan2018} .
\item{} 
The sample only includes galaxies with spectroscopic redshifts in the range of $0.05 < z < 0.3$.
Here, we excluded galaxies of $z < 0.05$ in order to remove misidentified 
stars and minimise the effect of peculiar motion to the conversion of redshift 
to  distance.  
We did not use the photo-z data, because, for low redshift galaxies, the photo-z is too uncertain. 
Galaxies in GOODS-North have spectroscopic redshifts from the 
Keck Treasury Redshift Survey \citep[TKRS][]{Wirth2004, Wirth2015}, which has a 
magnitude limit of r=24.4. So, the spectroscopy redshift survey limit is fainter
than our target selection limit in (1).
\item{} Edge-on galaxies of $b/a < 0.3$ were removed because they
may suffer from large extinctions, and therefore the UV and NIR as SFR and stellar mass tracers 
may be highly uncertain.
\item{} We performed visual checks on the HST images to make sure that the targets 
we selected are reliable. 
We also removed galaxies
with close neighbours (within 5 kpc), which would contaminate the radius measurement.
\end{itemize}

\begin{figure}[ht!]
\centering
\includegraphics[width=0.45\textwidth]{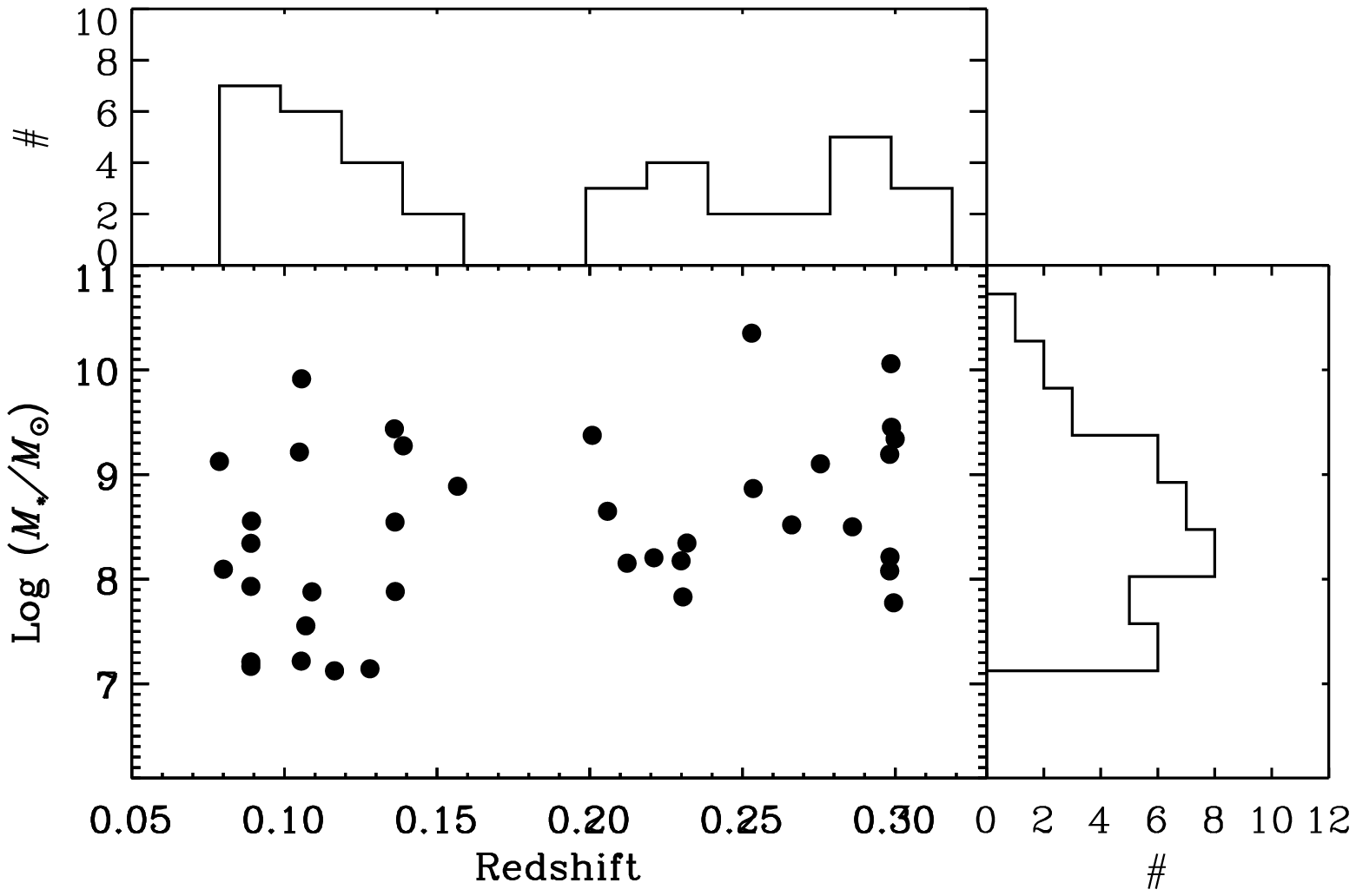}
\caption{Spec-z v.s. stellar mass of our low-redshift galaxy sample in the GOODS-North field. 
We also show the histogram of the  and redshift. The stellar masses are in the range of
$10^7 - 10^{10.5} M_\odot$.
}\label{hist-z}
\end{figure}

Our final sample contains 43 galaxies. The bias of the selection will be discussed in Section \ref{Discussion}.

\begin{table*}
        \centering
        \caption{Galaxy sample}
        \label{tab}
        \begin{tabular}{lccccccccccccc} 
                \hline
                \hline
                ID   & RA         & Dec        & $\log (M_{*}/M_{\odot})$ &  $z_{\rm spec}$  & $R_{\rm F275W}$ (kpc) & $R_{\rm F160W}$ (kpc) \\
                \hline            
        3141 & 12:36:52.3 & 62:09:31.9 & 8.17500                 &  0.2299 $^{(a)}$  &  1.74 $\pm$ 0.10 & 1.466 $\pm$   0.016 \\
        3238 & 12:36:50.0 & 62:09:35.5 & 8.50021                 &  0.2860 $^{(a)}$  &  1.87 $\pm$ 0.29 & 1.588 $\pm$   0.013 \\
        3384 & 12:36:51.1 & 62:09:38.6 & 8.64880                 &  0.2058 $^{(a)}$  &  0.69 $\pm$ 0.02 & 0.941 $\pm$   0.003 \\
        3800 & 12:36:51.6 & 62:09:54.5 & 8.54579                 &  0.1362 $^{(a)}$  &  0.36 $\pm$ 0.01 & 0.678 $\pm$   0.002 \\
        4739 & 12:37:04.3 & 62:10:29.9 & 8.07923                 &  0.2982 $^{(a)}$  &  0.59 $\pm$ 0.06 & 0.523 $\pm$   0.009 \\
        5784 & 12:37:17.5 & 62:10:57.0 & 9.01571                 &  0.2758 $^{(a)}$  &  $ <0.42$        & 0.435 $\pm$   0.002 \\
        7005 & 12:37:17.7 & 62:11:27.3 & 8.19215                 &  0.2131 $^{(a)}$  &  2.36 $\pm$ 0.21 & 2.223 $\pm$   0.019 \\
        7086 & 12:37:05.6 & 62:11:29.2 & 7.88190                 &  0.1363 $^{(a)}$  &  2.10 $\pm$ 0.38 & 1.657 $\pm$   0.015 \\
        7370 & 12:37:02.0 & 62:11:22.9 & 9.43692                 &  0.1360 $^{(a)}$  &  3.13 $\pm$ 0.07 & 3.181 $\pm$   0.002 \\
        7431 & 12:36:41.6 & 62:11:31.8 & 8.55460                 &  0.0892 $^{(a)}$  &  1.90 $\pm$ 0.02 & 1.372 $\pm$   0.002 \\
        7958 & 12:36:33.1 & 62:11:33.7 & 8.09484                 &  0.0800 $^{(b)}$  &  1.87 $\pm$ 0.21 & 1.625 $\pm$   0.007 \\
        8217 & 12:36:36.9 & 62:11:34.9 & 9.12485                 &  0.0787 $^{(c)}$  &  2.39 $\pm$ 0.02 & 2.365 $\pm$   0.001 \\
        8678 & 12:37:25.9 & 62:12:06.5 & 7.93125                 &  0.0890 $^{(b)}$  &  1.12 $\pm$ 0.13 & 1.350 $\pm$   0.007 \\
        9549 & 12:36:51.7 & 62:12:20.2 & 9.34111                 &  0.3000 $^{(a)}$  &  1.59 $\pm$ 0.18 & 1.270 $\pm$   0.002 \\
       10166 & 12:37:02.2 & 62:12:43.2 & 7.55415                 &  0.1070 $^{(b)}$  &  1.13 $\pm$ 0.21 & 1.055 $\pm$   0.008 \\
       10907 & 12:36:51.4 & 62:13:00.6 & 7.16645                 &  0.0890 $^{(b)}$  &  1.27 $\pm$ 0.42 & 1.514 $\pm$   0.030 \\
       11005 & 12:37:05.7 & 62:13:03.3 & 7.87935                 &  0.1090 $^{(b)}$  &  1.10 $\pm$ 0.18 & 1.112 $\pm$   0.005 \\
       12201 & 12:37:21.3 & 62:12:47.3 & 9.91500                 &  0.1056 $^{(a)}$  &  4.10 $\pm$ 0.13 & 3.664 $\pm$   0.002 \\
       12213 & 12:36:51.1 & 62:13:20.7 & 9.37483                 &  0.2008 $^{(a)}$  &  4.09 $\pm$ 0.12 & 3.688 $\pm$   0.004 \\
       12943 & 12:37:29.8 & 62:13:49.0 & 9.19334                 &  0.2982 $^{(a)}$  &  0.67 $\pm$ 0.11 & 0.867 $\pm$   0.003 \\
       13142 & 12:37:33.4 & 62:13:40.3 & 9.21471                 &  0.1049 $^{(a)}$  &  1.10 $\pm$ 0.02 & 1.702 $\pm$   0.002 \\
       13568 & 12:36:46.5 & 62:14:07.6 & 7.14365                 &  0.1280 $^{(b)}$  &  0.17 $\pm$ 0.12 & 0.517 $\pm$   0.010 \\
       13756 & 12:36:27.4 & 62:14:11.5 & 7.21730                 &  0.1055 $^{(a)}$  &  0.64 $\pm$ 0.35 & 0.817 $\pm$   0.013 \\
       13885 & 12:36:59.2 & 62:14:07.4 & 7.45163                 &  0.0887 $^{(a)}$  &  0.39 $\pm$ 0.63 & 1.523 $\pm$   0.042 \\
       14131 & 12:36:59.4 & 62:14:04.8 & 8.34265                 &  0.0890 $^{(b)}$  &  2.96 $\pm$ 0.29 & 2.333 $\pm$   0.008 \\
       14195 & 12:36:27.4 & 62:14:19.3 & 7.21000                 &  0.0890 $^{(b)}$  &  0.73 $\pm$ 0.46 & 1.114 $\pm$   0.017 \\
       14302 & 12:36:17.4 & 62:14:16.4 & 8.88805                 &  0.1567 $^{(a)}$  &  1.31 $\pm$ 0.02 & 1.108 $\pm$   0.002 \\
       14959 & 12:36:57.6 & 62:14:37.9 & 8.51865                 &  0.2661 $^{(a)}$  &  4.29 $\pm$ 0.78 & 3.429 $\pm$   0.033 \\
       14998 & 12:36:48.3 & 62:14:26.5 & 9.27473                 &  0.1389 $^{(a)}$  &  1.67 $\pm$ 0.01 & 1.406 $\pm$   0.001 \\
       15789 & 12:37:23.5 & 62:14:48.3 & 10.0798                 &  0.2533 $^{(a)}$  &  3.93 $\pm$ 0.05 & 3.351 $\pm$   0.002 \\
       16131 & 12:36:57.9 & 62:15:07.2 & 7.12445                 &  0.1164 $^{(a)}$  &  0.53 $\pm$ 0.08 & 0.644 $\pm$   0.009 \\
       16501 & 12:36:24.2 & 62:15:14.5 & 8.20445                 &  0.2210 $^{(a)}$  &  0.68 $\pm$ 0.08 & 0.652 $\pm$   0.005 \\
       17041 & 12:37:11.8 & 62:15:14.9 & 9.45124                 &  0.2988 $^{(a)}$  &  5.79 $\pm$ 0.14 & 4.468 $\pm$   0.007 \\
       18518 & 12:36:30.4 & 62:15:58.6 & 8.34530                 &  0.2318 $^{(a)}$  &  1.08 $\pm$ 0.10 & 1.219 $\pm$   0.007 \\
       20585 & 12:36:58.8 & 62:16:37.8 & 10.0591                 &  0.2986 $^{(a)}$  &  3.53 $\pm$ 0.14 & 3.544 $\pm$   0.002 \\
       22145 & 12:36:51.5 & 62:17:33.2 & 8.15263                 &  0.2122 $^{(a)}$  &  $ <0.35$        & 0.525 $\pm$   0.004 \\
       22453 & 12:37:06.5 & 62:17:29.1 & 10.3028                 &  0.2995 $^{(a)}$  &  7.16 $\pm$ 0.05 & 4.576 $\pm$   0.003 \\
       23354 & 12:36:52.8 & 62:18:07.7 & 10.3506                 &  0.2530 $^{(a)}$  &  7.84 $\pm$ 1.07 & 3.359 $\pm$   0.005 \\
       25910 & 12:36:41.1 & 62:18:55.3 & 8.21195                 &  0.2983 $^{(a)}$  &  1.24 $\pm$ 0.27 & 1.012 $\pm$   0.011 \\
       26119 & 12:36:49.6 & 62:19:16.3 & 9.10254                 &  0.2755 $^{(a)}$  &  5.12 $\pm$ 0.42 & 3.807 $\pm$   0.016 \\
       26772 & 12:36:46.0 & 62:19:10.3 & 7.77410                 &  0.2995 $^{(a)}$  &  $ <0.44$        & 1.111 $\pm$   0.024 \\
       27007 & 12:36:56.3 & 62:18:53.2 & 8.86610                 &  0.2535 $^{(a)}$  &  1.40 $\pm$ 0.20 & 1.538 $\pm$   0.006 \\
       27201 & 12:36:49.9 & 62:18:39.9 & 7.83055                 &  0.2305 $^{(a)}$  &  0.33 $\pm$ 0.03 & 0.525 $\pm$   0.007 \\
                \hline
        \end{tabular}
        
        Spec-z reference: (a), Wirth et al. 2004; (b): Reddy et al. 2006; (c): Cooper et al. 2011.
\end{table*}

\begin{figure}[ht!]
\centering
\includegraphics[width=0.5\textwidth]{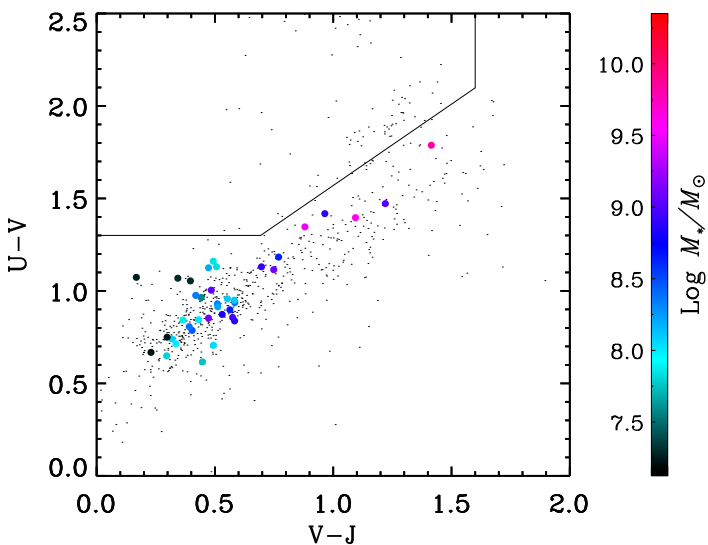}
\caption{The UVJ diagram of our sample. The black dots are the $0.05<z<0.3$ sample from CANDELS catalogue. 
This plot is colour-coded by the stellar mass.
All of the low-mass galaxies are in the star-forming region, and only the massive galaxies are dusty.} \label{MSUVJ}
\end{figure}

The redshift and mass histograms of our sample are presented in Fig. \ref{hist-z},
and the UVJ diagram of our sample is plotted in Fig. \ref{MSUVJ}. 
We use the median of the stellar mass measurement results \citep{Mobasher2015, Nayyeri2017},
which proved to be accurate for a wide range of redshift. 
Our sample covers wide ranges of stellar mass and colours, reaching the stellar mass of about $10^7 M_\odot$.
All of our sample galaxies have F275W detections.

\subsection{Radius Measurements}

We fixed
the galaxy centre from the HST WFC3 F160W image, then we measured the F160W, F275W image half-light radius from 
the dual image mode of SExtractor (v2.19), with the F160W image as the detection image, and the F275W image as
the measurement image. The PSF FWHM of the F275W and F160W images are 0.1 and 0.17 arcsec \citep{Oesch2018}.
We convolve the F275W image with a kernel with $\rm FWHM=\sqrt{\rm FWHM_{\rm F160W}^2 - \rm FWHM_{\rm F275W}^2}$, so that the 
F275W image would have a comparable spatial resolution. All of the spec-z selected galaxies have clear F275W and F160W 
detection (the flux/flux-errors given by SExtractor are lager than five), thus our sample all have reliable radius measurement.
Indeed, spec-z selected sample would bias to the UV bright galaxies. 
Three galaxies have a F275W radius lower than the image resolution: we denote their radii as upper limits that lower than the
PSF FWHM in Fig. \ref{radius}. 
We also de-convolve the measured half-light radius by $ \rm Half\,light\,radius = \sqrt{\rm Measured\,\,radius^2 - FWHM^2}$ \citep{Curtis2016}. 
This correction does not change our conclusion. We convert the radius into the unit of kpc based on the spec-z.

\section{Results}

\begin{figure}[ht!]
\centering
\includegraphics[width=0.6\textwidth]{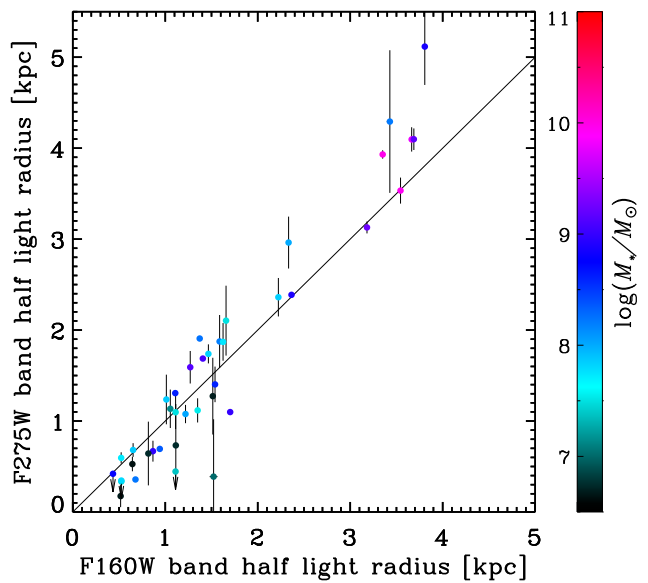}
\caption{F275W and F160W radius of our sample. For the large size galaxies, the F275W radius is typically larger 
than the F160W radius. For the galaxy with an average NUV radius lower than 1 kpc, the F275W radii 
are smaller, indicating a compact UV core in the galaxy.}\label{radius}
\end{figure}

We compare the F275W and F160W half-light radius in Fig. \ref{radius} and denote the stellar mass by the colour bar. 
For the galaxies with the NIR size larger than 2 kpc, their F160W radii are smaller than that of the F275W band.
{\rm The colour bar in Fig. \ref{radius} shows that the low-mass galaxies have a smaller F275W radius than the F160W radius.}

\begin{figure}[ht!]
\centering
\includegraphics[width=0.47\textwidth]{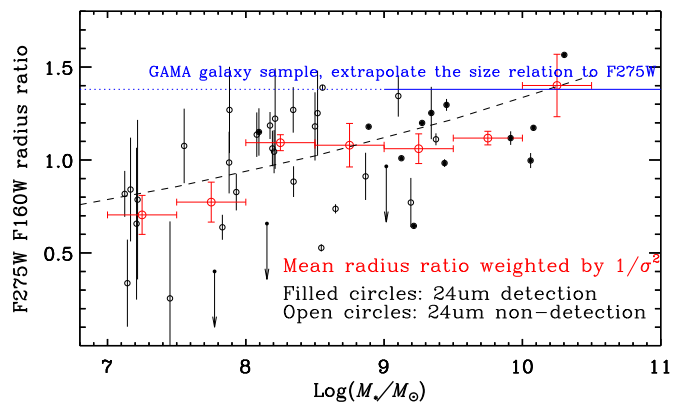}
\caption{F275W, F160W radius ratio and stellar mass for our sample.
Few galaxies have a high F275W to F160W radius ratio for the galaxy with the stellar mass lower than $10^8M_{\odot}$. 
The filled circles are the targets with 24$\mu$m detection, and the open circles are the targets that could not be detected 
in the 24$\mu$m image. The blue line shows the F275W to F160W size ratio that extrapolate from the results of the GAMA 
galaxy sample \citep{Kelvin2012}.  To better characterise the trend of the radius ratio as a function of the stellar mass, 
we liner fit the $\log (R_{\rm F275W}/ R_{\rm F160W})$ and the stellar mass and show the result with the dash line.
To show the radius ratio transition trend along the stellar mass, we also show the mean radius ratio (weighted by $1/\sigma^2$)  
in each 0.5 dex stellar mass bins with red open circles, and their error bars show the standard deviation of the mean.
}\label{UVIRratio}
\end{figure}

To further understand how the radius ratio changes with the stellar mass, we show the relation between the stellar mass
and the radius ratio in Fig. \ref{UVIRratio}. 
Though there is a larger scatter of the F275W and F160W radius ratio, we can see a clear trend
that the low-mass galaxies in our sample have lower radius ratios than the high mass galaxies.
To better characterise the transition, we show the mean radius ratios (weighted by $1/\sigma^2$) of 0.5 dex stellar mass
bins in red open circles. The average radius ratio drops from massive galaxies to the low-mass galaxies continuously. 
For galaxies with a stellar mass lower than about $10^{8} M_\odot$, the average radius ratio is 1$\sigma$ lower 
than the average ratios of galaxies with stellar mass higher than $10^{8} M_\odot$. 
But, as we are limited by the the sample size in each mass bin and the current radius ratio scatter, we cannot conclude a stellar-mass upper-limit of the `outside-in' growth mode.
We liner fit the $\log (R_{\rm F275W}/ R_{\rm F160W})$ and $M_{*,}$ and obtain a relation
$\log (R_{\rm F275W}/ R_{\rm F160W}) = -0.64 \pm 0.20 + (0.08 \pm 0.02) \times \log(M_{*}/M_\odot)$.

The massive galaxies are also bright and thus have high S/N in F275W and F160W images, so the radius measurement uncertainty
is smaller than the low-mass galaxies
Considering the 1$\sigma$ error bar of the radius ratio, the low-mass galaxies in our sample mainly have a smaller UV radius, 
hence an `outside-in' mode ,while galaxies in the massive end show an `inside-out' mode. 

Previous half light radius studies based on the Galaxy And Mass Assembly (GAMA) data sets showed that the radius changes 
with the wavelength \citep{Kelvin2012}: for images of the u, g, r, i, z bands from SDSS,
and the Y, J, H, Ks bands from UKIRT Infrared Deep Sky Survey Large Area Survey, the u band images of galaxies show the 
largest size (about 5.5 kpc), while the K band images show the smallest size (about 3.5 kpc). 
\citet{Kelvin2012} obtained a correlation between the wavelength and half light radius: $\log r_{\rm e, disk} = -0.189\log \lambda_{\rm rest} + 1.176$
\citet[see also][]{Evans1994, Cunow2001, Mollenhoff2006, Graham2008}.
According to this relation, we found that the F275W and F160W band radius ratio would be about  $5.2 \rm kpc / 3.7 kpc = 1.4$,
as shown in Fig. \ref{UVIRratio} with the blue line. It appears that the low-mass galaxies in our sample ($M_{*} \simeq 10^{7.5}M_\odot$)
have a systematically lower radius ratio than the massive galaxies (e.g. the GAMA galaxy sample).

\begin{figure}[ht!]
\centering
\includegraphics[width=0.5\textwidth]{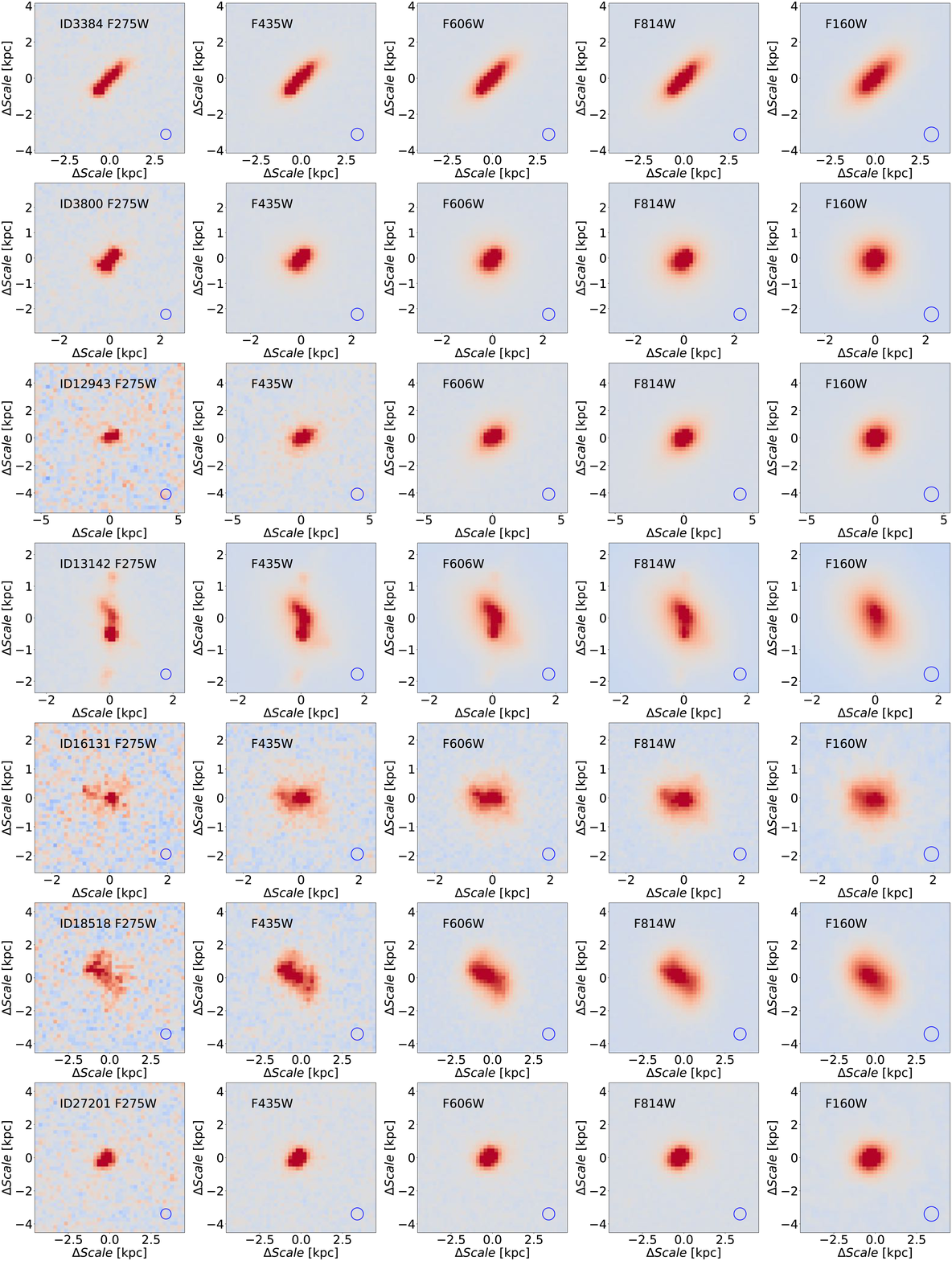}
\caption{Examples of UV compact galaxy sample. Each line shows one target in several bands.
We denote the CANDELS ID and bands name in the stamp images. The PSF of each image is shown by the blue circle.
The image scale has been transferred to the unit of kpc based on the spec-z of each target.
}\label{stamp}
\end{figure}

\begin{figure*}[ht!]
\centering
\includegraphics[width=0.47\textwidth]{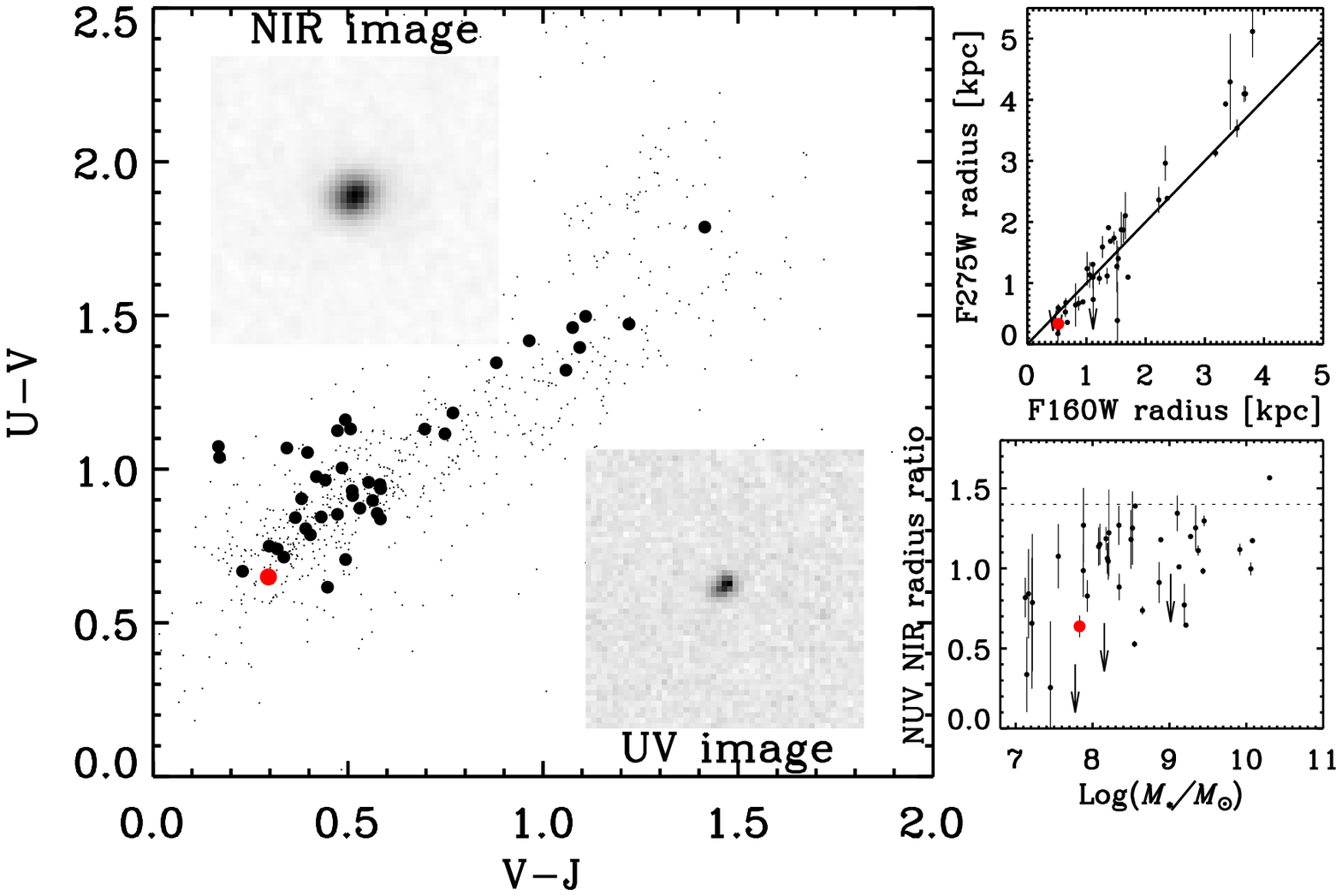}
\includegraphics[width=0.47\textwidth]{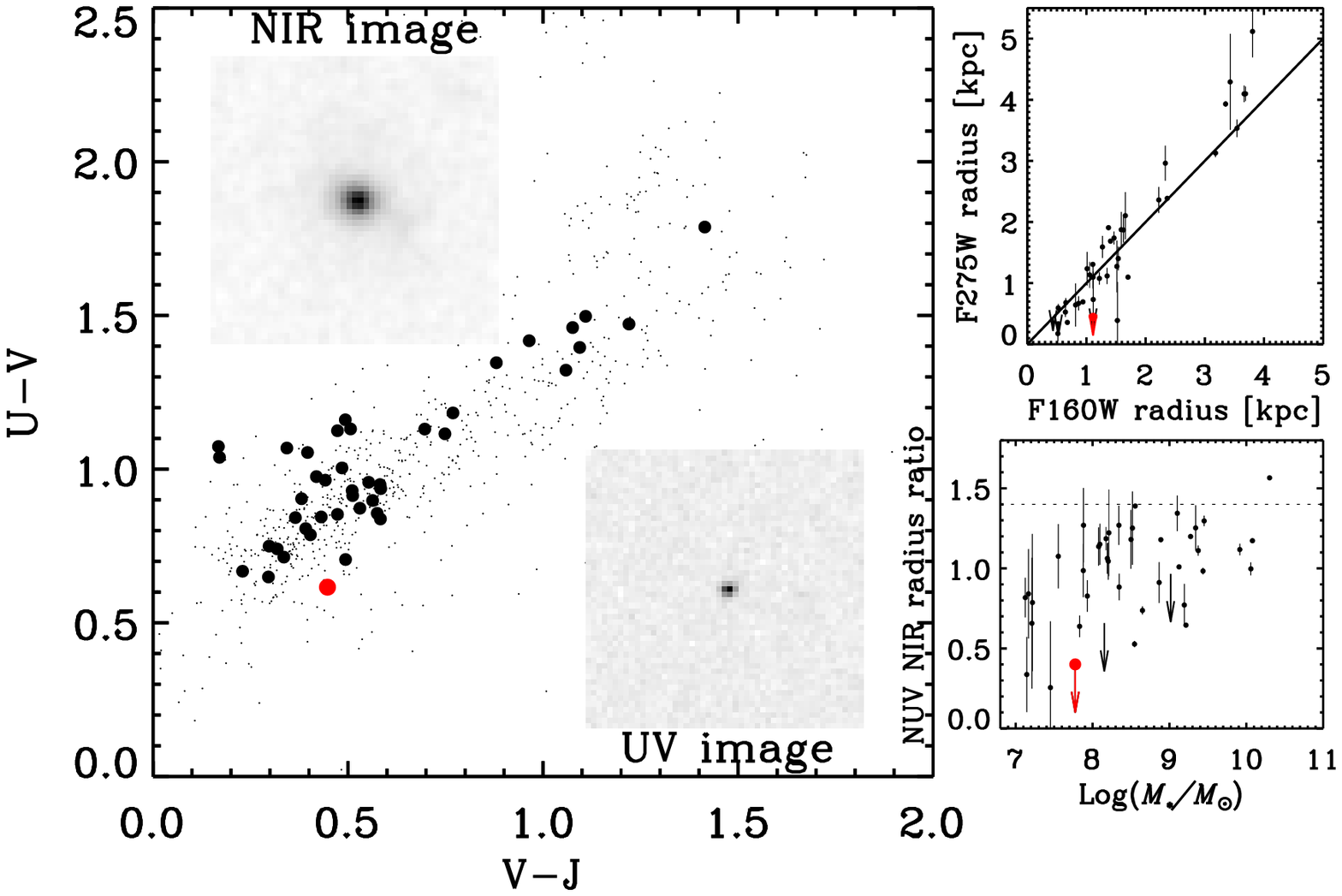}
\includegraphics[width=0.47\textwidth]{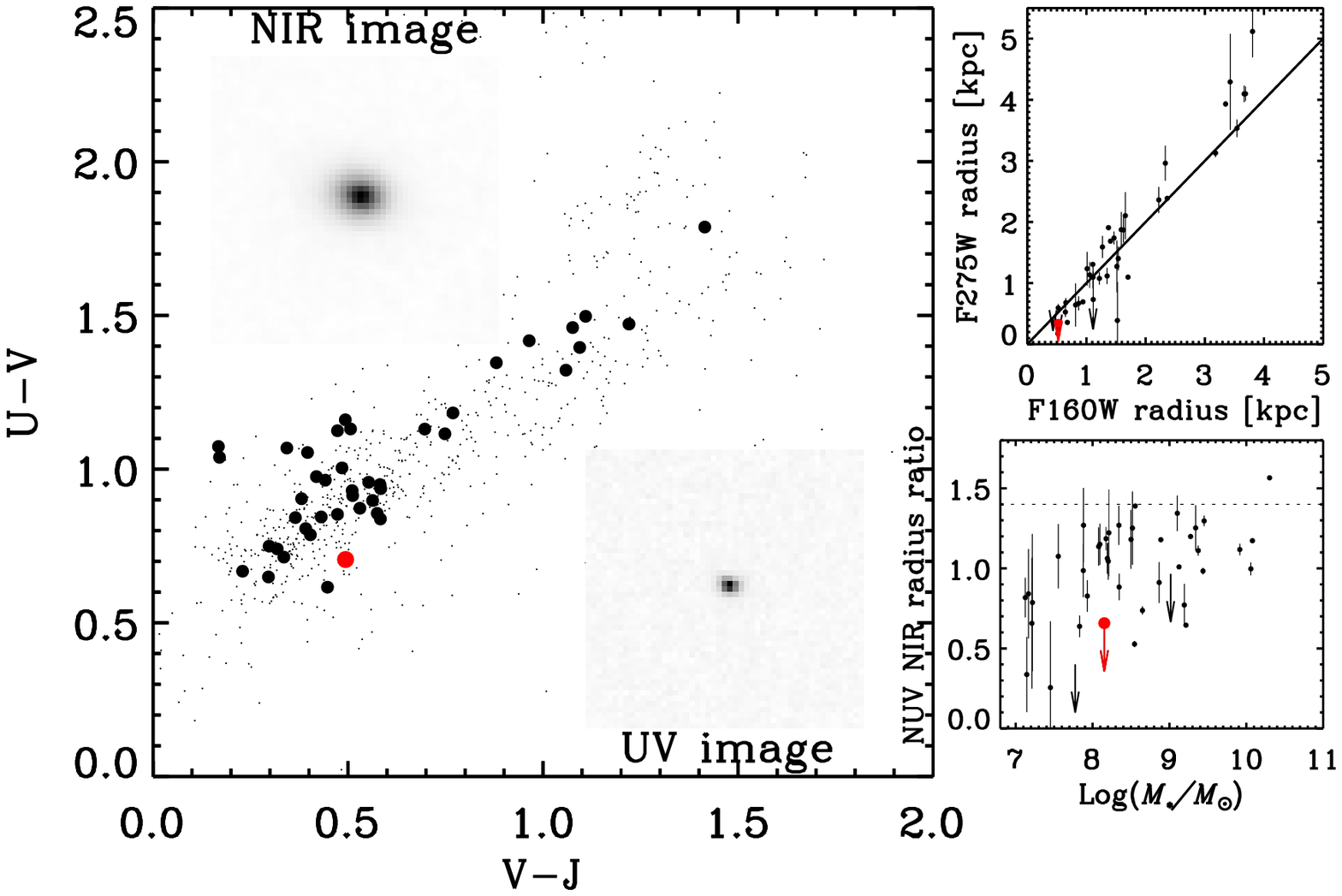}
\includegraphics[width=0.47\textwidth]{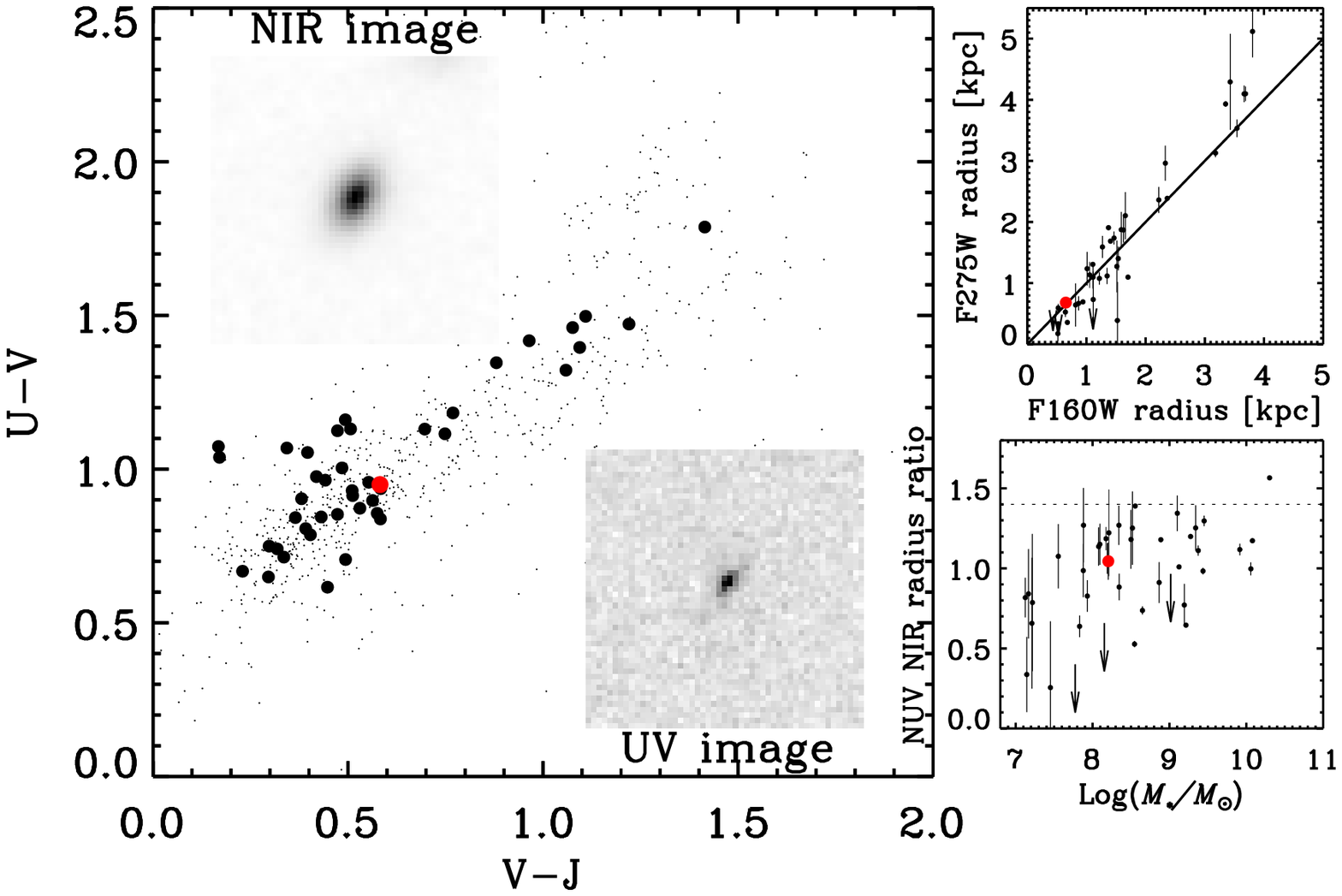}
\caption{Examples of low-mass galaxies. The two stamp images in each panel are the F160W and F275W images.
The red dots indicate the location of the example galaxy in UVJ diagram, F275W/F160W radius diagram, and the 
stellar mass v.s. the F275W/F160W radius ratio map. The small black dots are the targets with $z_{\rm spec}<0.5$ 
in CANDELS catalogue. We can see a very compact UV core in the galaxy centre, 
leading to a small star-formation region.
}\label{example_lowmass}
\end{figure*}

\begin{figure*}[ht!]
\centering
\includegraphics[width=0.47\textwidth]{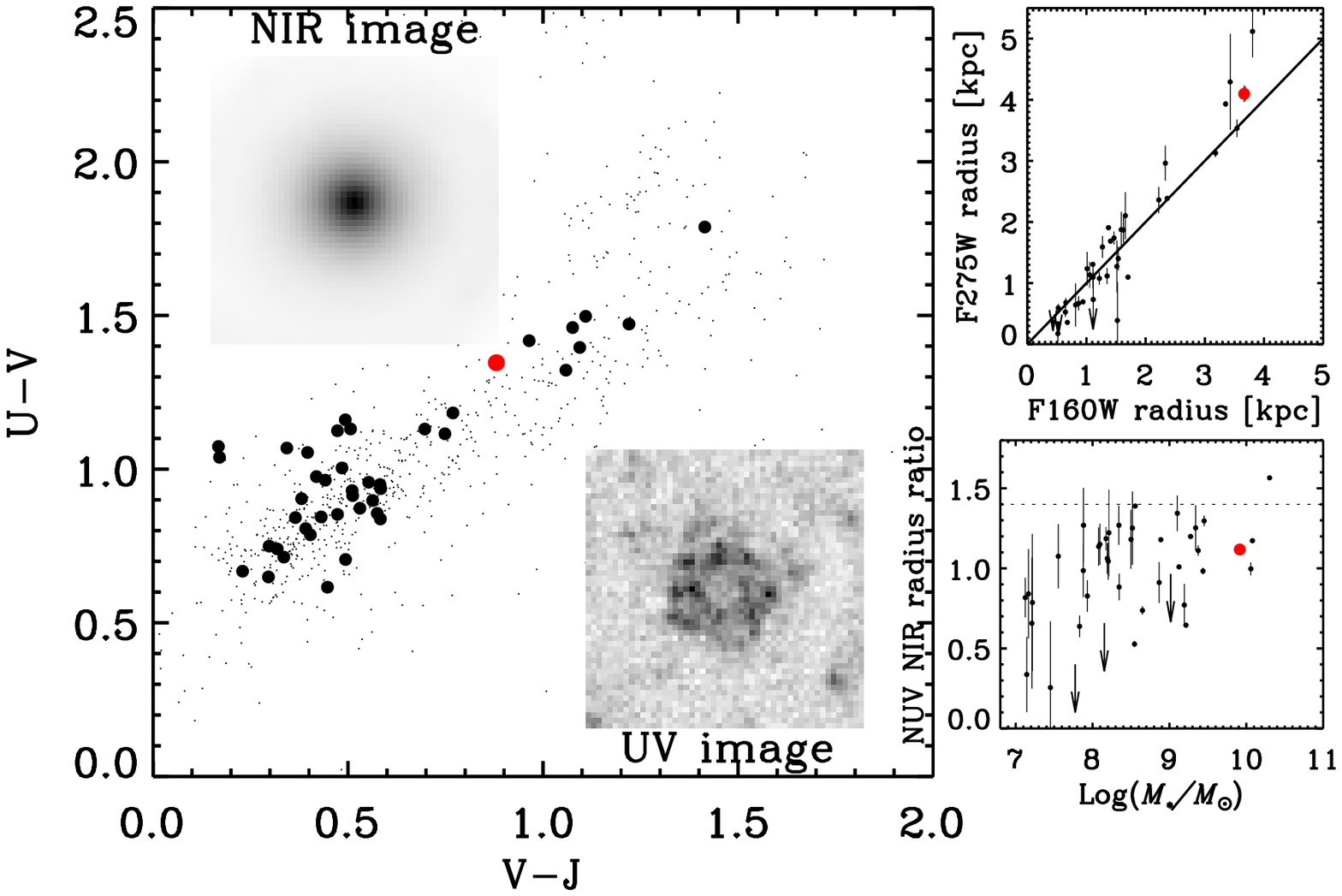}
\includegraphics[width=0.47\textwidth]{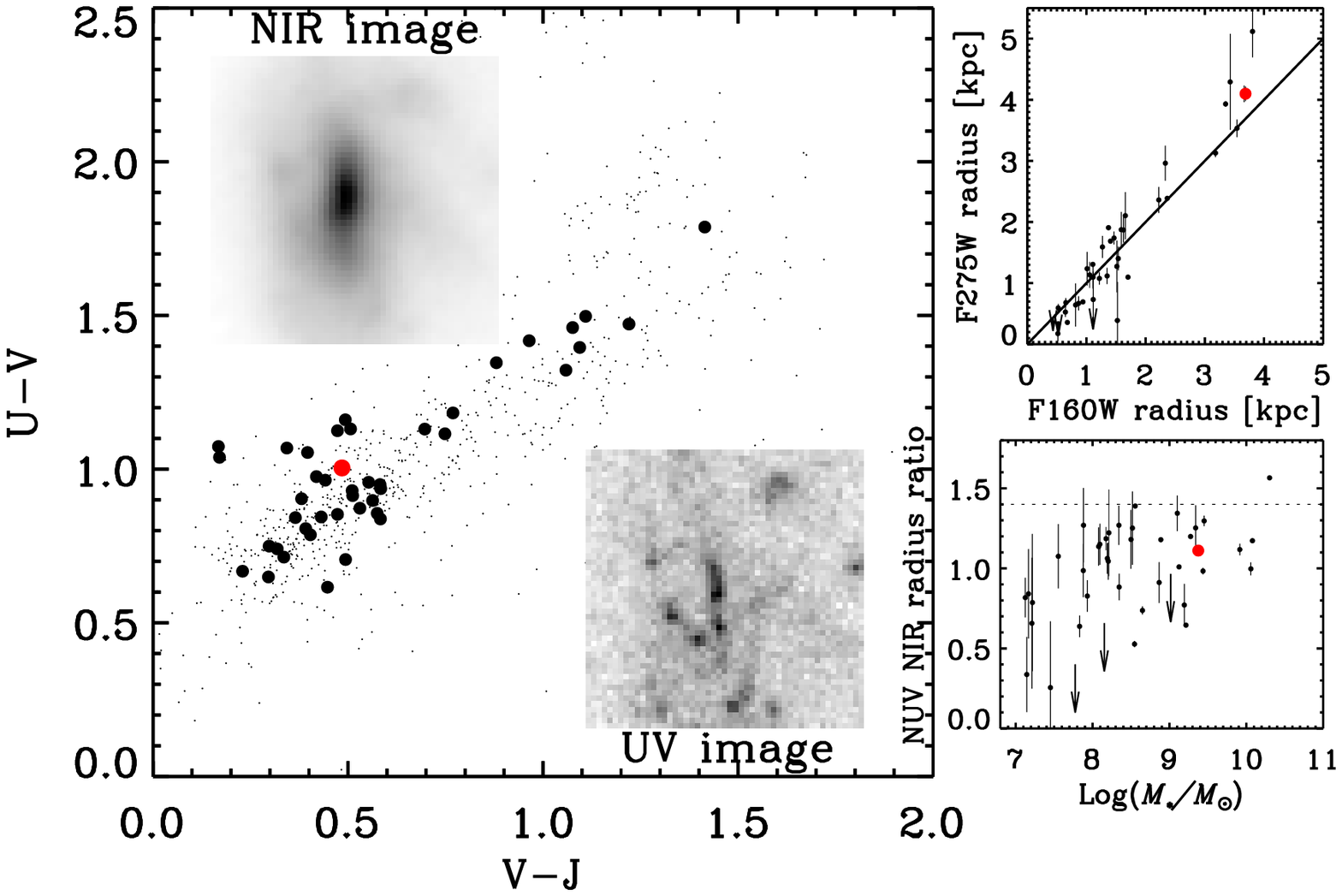}
\includegraphics[width=0.47\textwidth]{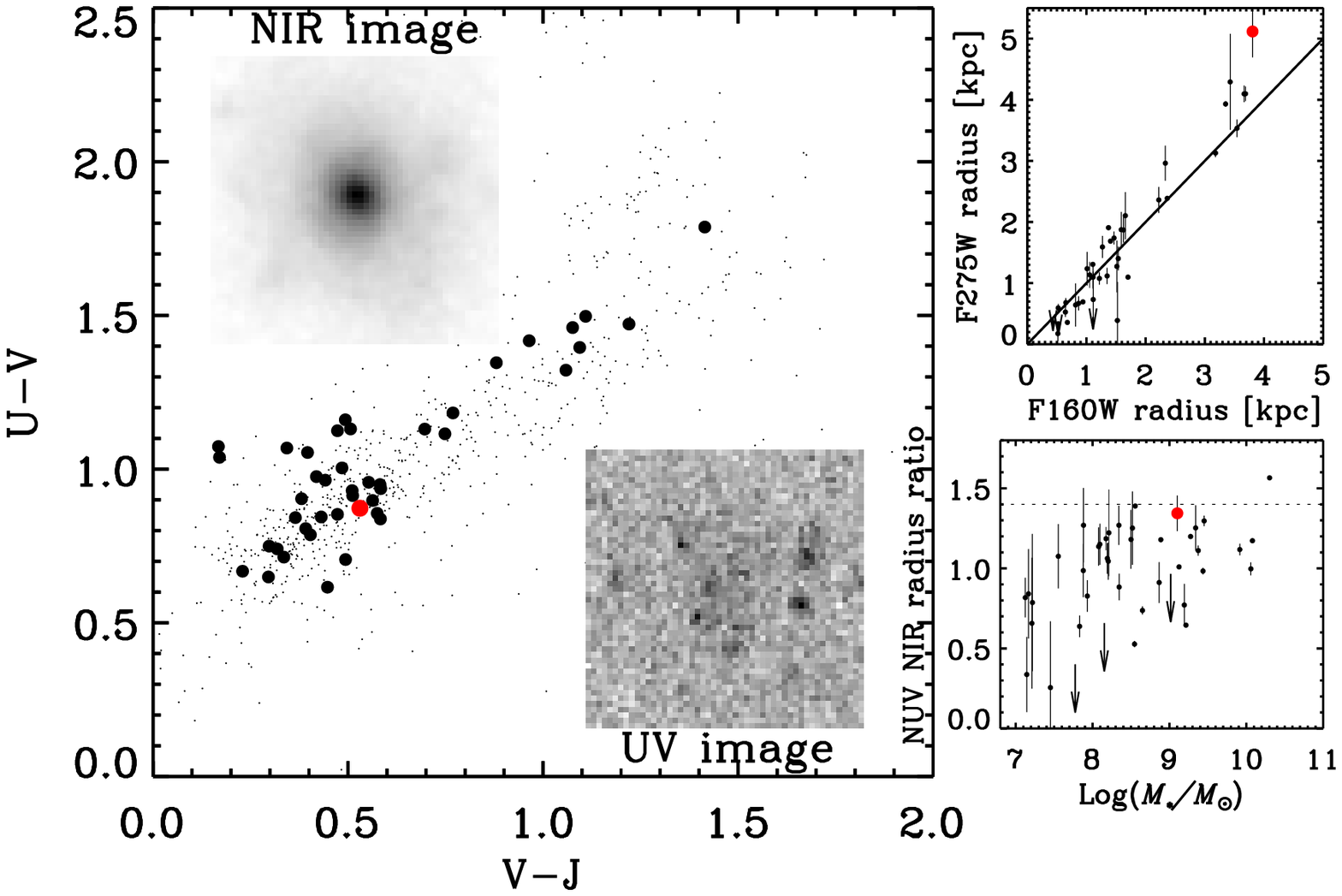}
\includegraphics[width=0.47\textwidth]{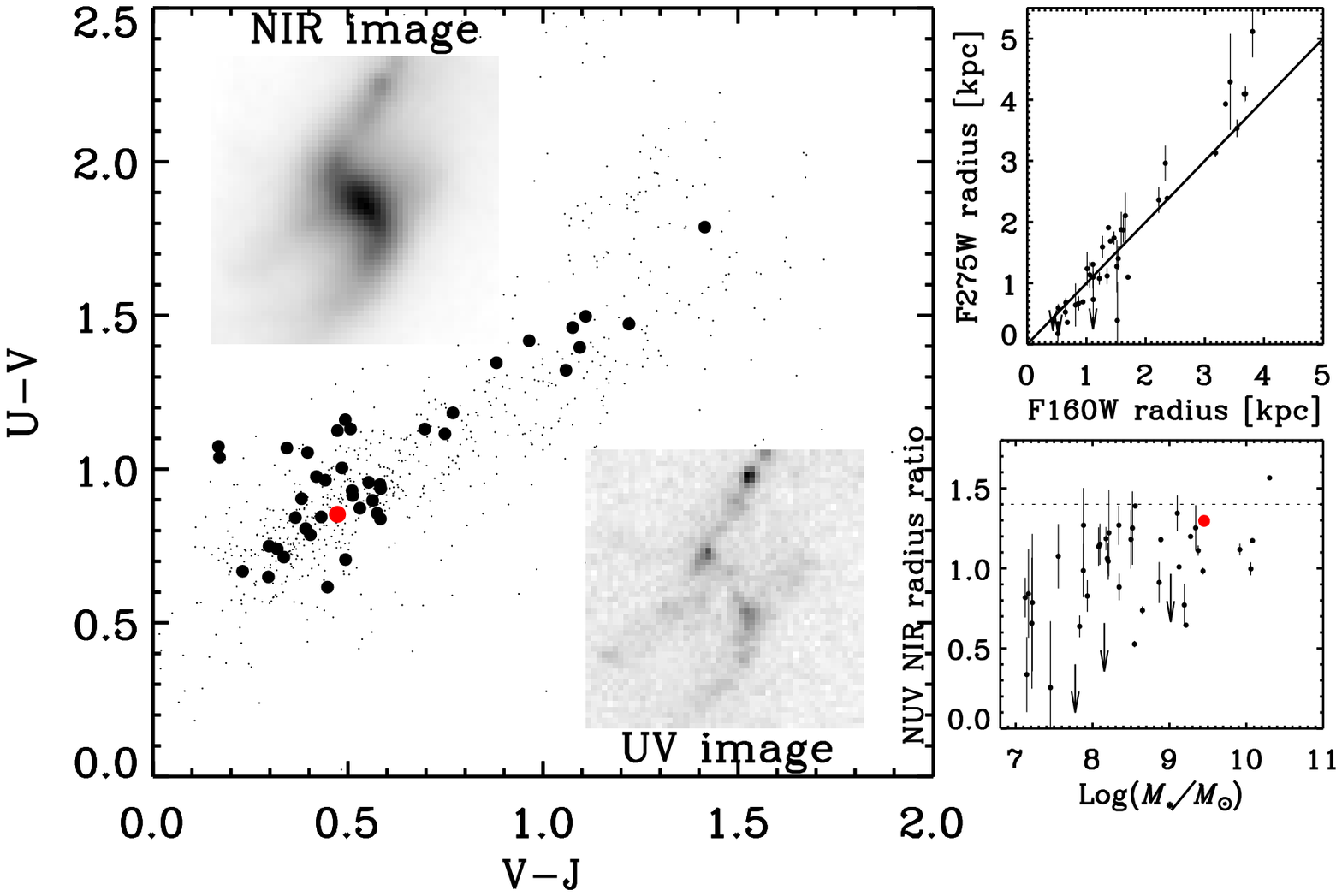}
\caption{Examples of high mass galaxies. The configuration of each panel is the same as Fig. \ref{example_lowmass}.
The UV morphology is more extended and more clumpy than the low-mass case,
indicating the ongoing unobscured star formations are in the disk, rather than the galaxy centre.
}\label{example_highmass}
\end{figure*}

We show seven galaxies as examples that have $r_{\rm F275W}/r_{\rm F160W} < 0.8$, with a stellar mass range from $7.1M_\odot$
to $9.2M_\odot$ and the median stellar mass $10^{7.8}M_\odot$.
Figure \ref{stamp} illustrates the UV compact galaxies in multi-bands. We list the F275W, F435W, F606W, F814W, F160W images
in each line for one galaxy. We show more comprehensive examples in Fig. \ref{example_lowmass}, \ref{example_highmass}.
In each panel, we show the stamp images in the F160W, F275W bands, and the red dots denote their location in the UVJ diagram, 
F275W v.s. F160W radius diagram and the stellar mass v.s. the F275W/F160W radius ratio diagrams. For the massive galaxies, 
we can that see most of the galaxies have clumpy UV morphology, which means some of the star-forming clumps are not quite suffered 
from the dust extinction.
For most of the low-mass galaxies, the UV morphology turns to be very compact. The compact star-formation core is very 
similar to the compact star-formation core in the ULIRGs \citep{Ma2015} and we can expect the compact star formation in these 
low-mass galaxies would build an old population bulge in the galaxy centre.

We also showed how the dust affects our sample in Fig. \ref{UVIRratio} by denoting the 24 $\mu$m detected targets 
as filled circles, and the 24 $\mu$m non-detected galaxies as open circles. 
The resolution of the Spitzer/MIPS 24um image is about 6'', much larger than that of the HST resolution. 
To avoid the blending issue, we match the catalogue 
from the Spitzer/MIPS 24$\mu$m catalogue (beam size 6'') to the 8$\mu$m catalogue (2" resolution), then match 
the 8$\mu$m counterparts from the CANDELS F160W band selected catalogue, so that one 24$\mu$m target will have at most one 
optical counterpart. For the catalogue-matching process, we consider the corrected-Poissonian probability, 
or the p-values \citep{Downes1986}. We can see that nearly all the galaxies with a stellar mass lower than 
$10^{8}M_{\odot}$ are not detected in the 24$\mu$m image.

The smaller F160W radius indicates the stars concentrate more in the galaxy centre, while the compact UV
morphology means the stars are forming in the galaxy centre.
Our results show that, for the galaxies with the stellar mass larger than $10^{8}M_\odot$, 
the stars are forming in the outskirt of the galaxy. This is consistent with the recent works of 
the comparison between the H$\alpha$ and NIR radius  for the $10^9<M_*/M_\odot<10^{11}$ galaxies at $0.7<z<1.5$ \citep[][]{Nelson2016, Nelson2019}, implying the `inside-out' 
scenario. Moreover, for the small-size galaxies, the UV morphology of some low-mass galaxies are more compact 
than their NIR image. While the star formation is still ongoing in the galaxy centre, 
the stars are already formed in the outer range of the galaxies.

\section{Discussion   }\label{Discussion}
We briefly discuss the possible selection bias, the galaxy growth mode, the environment affection, 
and the possible origin of the `outside-in' growth mode in this section.

\subsection{Selection bias}

Our results show the existence of the UV compact galaxy in the low redshift low-mass galaxy populations,
indicating an `outside-in' mode of stellar assembly.
However, our results may be biased by the selection methods. Here, we discuss the possible affection to our results
of the selection bias.

\subsubsection{Sample selection}

Here we only select the low redshift ($0.05 < z < 0.30$) galaxies in order to study the high-resolution rest frame near ultraviolet (NUV) and NIR HST 
images simultaneously.
Our sample does not include galaxies without spectroscopically measured redshifts. 
However, this does not introduce any significant bias to our main results.
The UV compact galaxies in our sample should belong to the blue galaxy population, in which it is easier to identify the redshifts.
Since the spec-z survey depth is deeper than our F606W band selection limit, the galaxies with F606W magnitude brighter than 24 AB mag
but have no spec-z results should not be the majority population, and thus not change our results,  
especially for our purpose of showing the existence of the UV compact low-mass galaxies.

\subsubsection{Dust extinction}

Dust in galaxies may affect the intrinsic UV scale measurement in that the dust may split the star-formation 
region into a clumpy morphology or even a UV faint image. 
Previous galaxy size given by \citet{Kelvin2012} showed the trend of anti-correlation between the SDSS -selected galaxy half light radius and the wavelength, which is explained by the dust extinction in a different band
\citep{Evans1994, Cunow2001, Mollenhoff2006, Graham2008}.

Observation results showed that, for low-mass galaxies, the dust and stellar-mass 
ratio are lower than the massive galaxies, so dusty low-mass galaxies are not the majority 
in the low-mass galaxy population \citep[e.g. the low-mass end of Fig. 4 in ][]{Fisher2014}. 
Dust is produced in some stellar processes \citep[e.g. the SNe or asymptotic giant branch star wind,][]{Sadavoy2019}.
For the low-mass galaxies, the stellar-mass-building history is still short. Therefore we
may expect that the dust in the low-mass galaxies still follows the stellar distribution, and is not likely to
affect the UV morphology by a ring-like dusty structure. 

Limited by the 1'' resolution of the SDSS image and survey depth, the results of \citet{Kelvin2012} are mainly
based on galaxies of $M_{*}>10^{9}M_\odot$. In contrast, our sample selected from the CANDELS catalogue has
the stellar mass range from $10^7 - 10^{10.5} M_{\odot}$ with high spatial resolution, which completes the
SDSS sample to a lower mass limit. The 24$\mu$m detection results in Fig. \ref{UVIRratio} show that low-mass
galaxies are not as dusty as the massive galaxies. The UVJ diagram in Fig. \ref{MSUVJ} also shows that few 
low-mass galaxies are in the dusty region. So, we conclude that the UV compact morphology for our sample
is not significantly affected by the dust extinction.

\subsubsection{AGN}

\begin{figure}[ht!]
\centering
\includegraphics[width=0.45\textwidth]{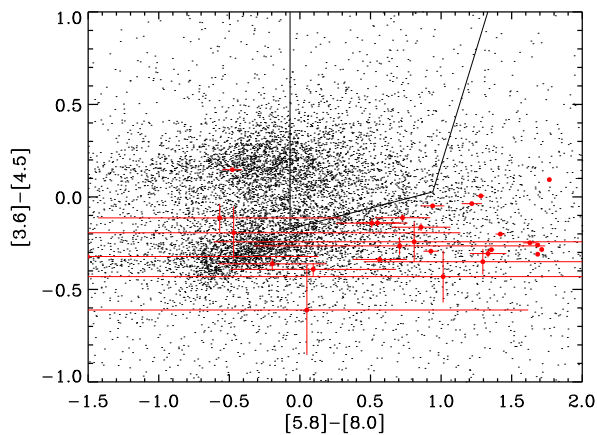}
\caption{Spitzer/IRAC colour-colour diagram of the GOODS-N sample and our low-redshift sample (red dots). All of our sample galaxies are not MIR power-law AGN. The large error bars in the [5.8]-[8.0] colour is caused by the low sensitivity of the 5.8$\mu$m image data. The low [3.6]-[4.5] colour indicates that our low-mass galaxy sample
contains very few AGNs. The black dots are the galaxies in CANDELS GOODS-North field as a comparison.}\label{irac}
\end{figure}

The existence of the AGN in the galaxy centre will produce a UV bright core and contaminate our results \citep{Dashyan2018}. 
For all galaxies in our sample, we checked the emission lines and found no broad emission lines. 
Moreover, since the AGN may heat the dust
and produce a `power-law' SED in the MIR bands, we also showed the Spitzer/IRAC colour-colour diagram \citep{Stern2005} in
Fig. \ref{irac}. AGNs would be located in the wedge region, and we can see that all our spec-z selected sample
is not in the AGN region. The colour uncertainties are mainly contributed by the [5.8]-[8.0], which is caused by the low-sensitivity of the 5.8$\mu$m filter. Nevertheless, the low [3.6]-[4.5] colour indicates that our sample should not contain much AGN.

\subsection{Transition from `outside-in' to `inside-out'}

Low-mass galaxies are found to have an `outside-in' growth mode, while the massive galaxies
commonly have `inside-out' modes \citep{Wang2011, Zhang2012}. However, how and when the growth mode
changes is still not quite clear. Results of Local Irregulars That Trace Luminosity Extremes-The HI Nearby Galaxy Survey 
\citep[LITTLE-THINGS,][]{Hunter2012}
show that the `outside-in' growth mode would happen for the galaxies with $M_*<10^8M_{\odot}$.
On the other hand, based on the spatially resolved galaxy colour from GALEX and SDSS image data, 
\citet{Pan2015} found that the `outside-in' growth mode should occur at a stellar mass lower $10^{10} M_\odot$. 
However, limited by the survey depth, SDSS can only provide a galaxy sample with $M_{*}>10^9M_{\odot}$, and
the low-mass galaxies in SDSS are mainly satellite galaxies \citep{Geha2012}, which may have a different quenching mode
from our sample (see the discussion in Sec. 4.3).

The HDUV and CANDELS projects provide a unique chance to study the galaxy with 
stellar mass range about $10^{7-11}M_{\odot}$ with high spatial resolution. And our result shows that galaxies
with stellar mass about $10^{7.5}M_{\odot}$ would have a `outside-in' mode. Moreover, Fig. \ref{UVIRratio} shows 
that there is a lack of low-mass galaxies ($\sim 10^{7.5}M_{\odot}$) with high radius ratios,
and a lack of massive galaxies (e.g. $> 10^{9}M_{\odot}$) with compact UV radii. 
This indicates that the transition from `outside-in' to `inside-out' growth mode
does not occur at a sharp mass criteria, but in a range of low stellar mass from $10^{7.5}M_{\odot}$ to 
$10^{9} M_\sun$, although this upper boundary of the mass range is poorly constrained, because measurement of the 
UV radii of more massive galaxies is increasingly affected by severe dust extinction.

\subsection{Environment}

The evolution path of the low-mass galaxies would be more easily affected by the environment. 
The tidal distance $d_{\rm t} \simeq r_{\rm galaxy} (3M/m) ^{1/3}$, where $r_{\rm galaxy}$ is the radius of the low-mass galaxy, 
$m$ and $M$ is the total mass of the low and high galaxies, respectively \citep{Binney2008}. For the two galaxies with the
dark matter halo ratio about 1000, the low-mass galaxy with a typical halo radius of about 10 kpc will lead to a tidal 
distance about 100 kpc.
Thus, a low-mass galaxy might be tidally affected by a 1000 times more
massive galaxy within 100 kpc scale, 
much larger than the two galaxies with similar mass.

Tidal from the nearby massive galaxy halo may remove the gas and quench the star-formation process \citep{Fang2016, Zinger2018}.
To understand the environment of our sample, especially whether the UV compact galaxies are 
field galaxies or satellite galaxies, we still need to identify the massive neighbour of our sample galaxy. 
The SDSS project with a large survey area and complete spec-z deep to r=17.7 provides the perfect 
massive galaxy sample to search the possible neighbour galaxy of our sample. 
If one of our sample galaxies has no SDSS galaxy at a similar redshift bin ($\Delta v < 500 \rm km/s$), 
we can conclude that this sample galaxy is not likely to be tidally affected by the nearby massive
galaxy.

\begin{figure}[ht!]
\centering
\includegraphics[width=0.45\textwidth]{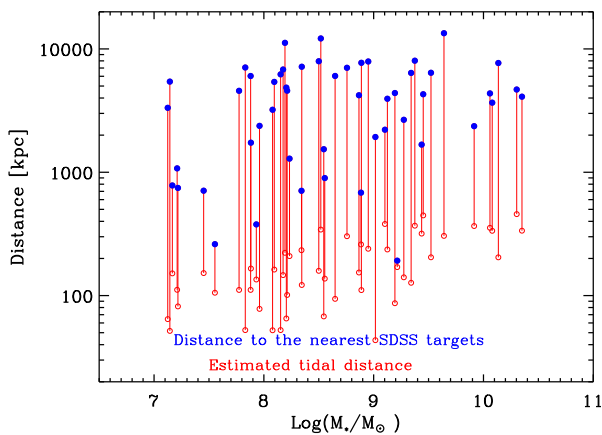}
\caption{Nearest distance between our galaxy sample and the SDSS galaxies. We search the SDSS galaxies with the velocity range from -500 to 500 km/s.
We also roughly estimate the distance that may tidally affect our galaxy sample and show the results by the open circles. All of the nearest SDSS
galaxies are still beyond the tidally affected distance. So, we conclude the origin of the `inside-out' or `outside-in' growth mode is more likely a result of secular evolution rather than the affection of the environment.}\label{distance}
\end{figure}

For each galaxy in our sample, we search for neighbour galaxies from SDSS dr12 data that satisfy: 
1, velocity range within [-500, 500] km/s; 
2, spatial distance within 1 degree, corresponding to a distance about 3.5 to 16 Mpc at the redshift $0.05 < z < 0.3$.
We show the nearest spatial distance in Fig. \ref{distance}. 
We can see most of the nearest SDSS galaxies are beyond 0.5 Mpc, which is the typical scale of the
galaxy cluster. This is consistent with the fact that no low-redshift galaxy cluster was reported close to GOODS-N.
So we conclude that most of the galaxies in our sample are not in the group or cluster environment. 

{The typical  }distance between our sample to the nearest SDSS galaxies is about 4 Mpc. So, the density is about
$4\times 10^{-3} \rm Mpc^{-3}$, corresponding to that of a typical galaxy of $10^{10-11} M_\odot$
\citep{Baldry2008}. Thus the distances shown in Fig. \ref{distance} indicate that the closest SDSS galaxy to our sample
is indeed massive galaxies, and the typical number density of the massive galaxy is too low to affect our sample.
Indeed, for a low-mass galaxy of mass $m$ with a companion of mass $M$ at distance D, the tidal radius is 
$r_t = D\times (m/3M)^{1/3}$. For $D \sim 4$ Mpc and $m/3M \sim 10^{-5}$ \citep[based on the stellar mass function
and the ratio between the dark matter and stellar mass in, for example, ][]{Baldry2008, Behroozi2013a, Behroozi2013b},  
$r_t \sim 100$ kpc. Low-mass galaxies in our sample have sizes far smaller than this $r_t$, therefore will be little affected by the neighbour galaxies.

Since the low-mass galaxies are more easily affected by the nearby field galaxies, we estimate the distance that our sample might be tidally affected by.
The mass ratio between the massive galaxy and the low-mass galaxy would be in the range of $10^3-10^6$, so term $(3M/m) ^{1/3}$ is roughly 10 to 100.
Then, the tidal distance that affects the galaxy at about mass scale (NIR image scale) or galaxy star-formation scale (about the UV image scale) should
be less than 100 times the galaxy scale. We also show this roughly estimated tidal distance in Fig. \ref{distance} with open circles. So, most of the massive galaxies
from SDSS are still beyond the tidal distance. We conclude that almost all of the spec-z selected galaxies are field galaxies. 
The origin of the UV compact morphology is more likely a secular evolution result, and not quite influenced by the galaxy environment.

\subsection{Origin and evolution of the UV compact galaxy}
\begin{figure}[ht!]
\centering
\includegraphics[width=0.6\textwidth]{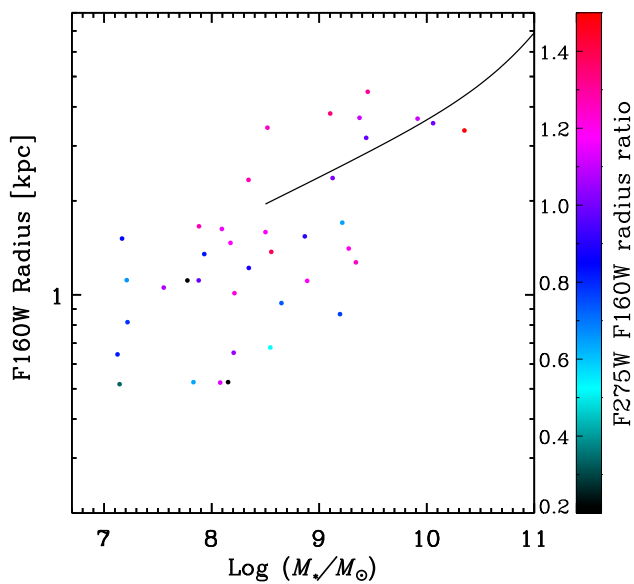}
\caption{Mass-size relation of our sample. The thick line shows the mass-size relation for star-forming galaxies 
in \citet{Lange2015} from the GAMA survey where the mass limit is about $10^{9.}$. Our sample is in agreement with 
the mass-size relation at $10^{9-10}M_\odot$, and more scatter at the low-mass end.}\label{massradius}
\end{figure}

Simulation suggests a star formation compaction process when the gas is more efficient in falling to the galaxy 
centre than the star formation\citep{Tacchella2016, Tacchella2018}, yielding a `blue nugget' phase.
The collision of the in-falling gas from a different direction would increase the gas density and instabilities
to trigger the starburst. In this case, the stars are mainly formed in the galaxy centre, which also
`shrinks' the stellar mass. Fig. \ref{massradius} shows the stellar mass v.s. the F160W size of our sample.
We also show the mass-size relation from the GAMA result \citep{Lange2015}, which is valid to the low-mass end
about $10^{8.5}M_\odot$. Galaxies with a stellar mass larger than $10^{9}M_\odot$ have a good consistency
with the previous relation. However, for the low-mass galaxies in our sample, the galaxy size has offset the 
trend of the massive end. The smaller size of the low-mass galaxies may be caused by the centre star formation, 
which will produce more stars in the galaxy centre. The large scatter of the mass-size relation for
the low-mass galaxies also indicates a different galaxy growth mode at low-mass end.

Stellar feedback processes such as SNe can produce galactic winds and remove the gas from the dark matter
potential. For the low-mass galaxies, the stellar feedback would be more efficient because of the shallow 
gravitational potential, and the outer region of the galaxy gas would be blown away, leaving only the gas 
in the central region. Thus the low-mass galaxies are more likely to have compact blue cores than massive galaxies\citep{Schawinski2009}.

\citet{Lin2019} show that it is the nearby dark matter halo mass, rather than the 
stellar mass, that affects the fraction of the `inside-out' quenching fraction, and the `outside-in' quenching
could be the result of the environment. Since we found that the SDSS Spec-z galaxies, 
which are mainly the stellar massive galaxies, are still too far to affect our sample, 
less-massive halos close to our sample may account for the formation mechanism of 
the UV compact galaxies we present here. 

The GALEX UV image survey has identified a rare population of the compact UV-luminous galaxies (UVLGs,
stellar mass about $10^{10}M_{\odot}$),
which are suggested as the low-redshift analogues of the Lyman-Break galaxies (LBGs) populations \citep{Heckman2005, Hoopes2007, Overzier2008, Overzier2010}.
The UV compact galaxies revealed by the HST images may also belong to a similar population to the compact UVLGs,
but have lower stellar mass and higher spatial resolution. The `outside-in' trend revealed from the comparison of
the UV and NIR radius also implies the high-redshift LBGs may have `outside-in' formation mode. 

Previous studies of the blue spheroid (BSph) galaxies \citep{Kannappan2009, Schawinski2009, Mahajan2018, Moffett2019} 
or Luminous Compact Blue (LCB) galaxies \citep{Noeske2006} revealed a population of the blue compact galaxies that 
found between $z=0$ and $z=1$, which are the candidates of the low-mass quiescent galaxy progenitor. 
This blue compact galaxy population is very likely to overlap with the UV compact galaxies we found here
directly from the HDUV image. Further study and comparison with the UV compact galaxies is to come in a forthcoming paper.

\section{Conclusion}

Using the recently released HDUV data and the CANDELS data of the GOODS-North field, we study the UV and NIR scale of the low-redshift galaxies to identify the star formation and stellar mass distributions. We find some of the low-mass galaxies ($<10^{8}M_\odot$, see Fig. 4)
have a very compact UV core in the galaxy centre. Further study shows that the UV compact morphology is not likely to be caused
by the dust extinction or AGNs. The smaller UV size indicates a star formation in the galaxy centre, while the stars are formed at
a larger radius, which supports the `outside-in' growth mode for low-mass galaxies. For the massive galaxies, our results also show 
an `inside-out' mode, which is consistent with the previous study. 
Our result show that the `outside-in' and `inside-out' growth modes transit smoothly with large scatter.

With the help of the SDSS data, we find such UV compact galaxies are field galaxies. 
The mass-size relation of our sample shows a smaller mass for the low-mass galaxies, 
which may be caused by the centre star formation in the low-mass galaxies.

The mass-size relation of our sample is consistent with the previous GAMA survey result for the galaxies with stellar mass $>10^9M_\odot$,
but more scatter for low-mass galaxies, indicating a different growth mode for low- and high-mass galaxies.

\begin{acknowledgements}
        We thank the referee for carefully reading and for providing constructive comments that helped us to improve the quality of this paper.
        C.C. would like to thank Prof. Hongxin Zhang, Prof. Xianzhong Zheng and Dr. Zhiyuan Ma for insightful comments and helpful discussions. 
		This work is sponsored (in part) by the Chinese Academy of Sciences (CAS), through a grant to the CAS South America Center for Astronomy (CASSACA) in Santiago, Chile.
C.C is supported by the National Natural Science Foundation of China (NSFC), No. 11803044 and the Young Researcher Grant of National Astronomical Observatories, CAS. W.D. is supported by the NSFC No. U1931109. Z.P. acknowledges the support from NSFC, No. 11703092.
R.L. acknowledges support by CONICYT-PIAACT No. 172033 and CONICYT and QUIMAL No. 160012. P.A. acknowledges the support from CAS-Conicyt 15020, the support provided by Fondecyt reg. n. 1170518, and Chilean BASAL Centro de Excelencia en Astrofisica y 
Tecnologias Afines (CATA) grant PFB-06/2007.
\end{acknowledgements}


\end{document}